\documentclass[final,times,twocolumn,authoryear]{elsarticle}
\usepackage[margin=2cm]{geometry}% by courtesy of Mico
            
\usepackage{amssymb}
\usepackage{array}

\usepackage{listings}

\usepackage{graphicx}
\usepackage{placeins}

\usepackage{xpatch}
\usepackage[font=normalfont]{subfig}

\usepackage{etoolbox,lipsum}
\makeatletter
\patchcmd\@maketitle\@author{\@author \\[\normalbaselineskip] \myfigure}{}{}
\newcommand\myfigure{%
  \makebox[0pt]{
 \includegraphics[height=3cm]{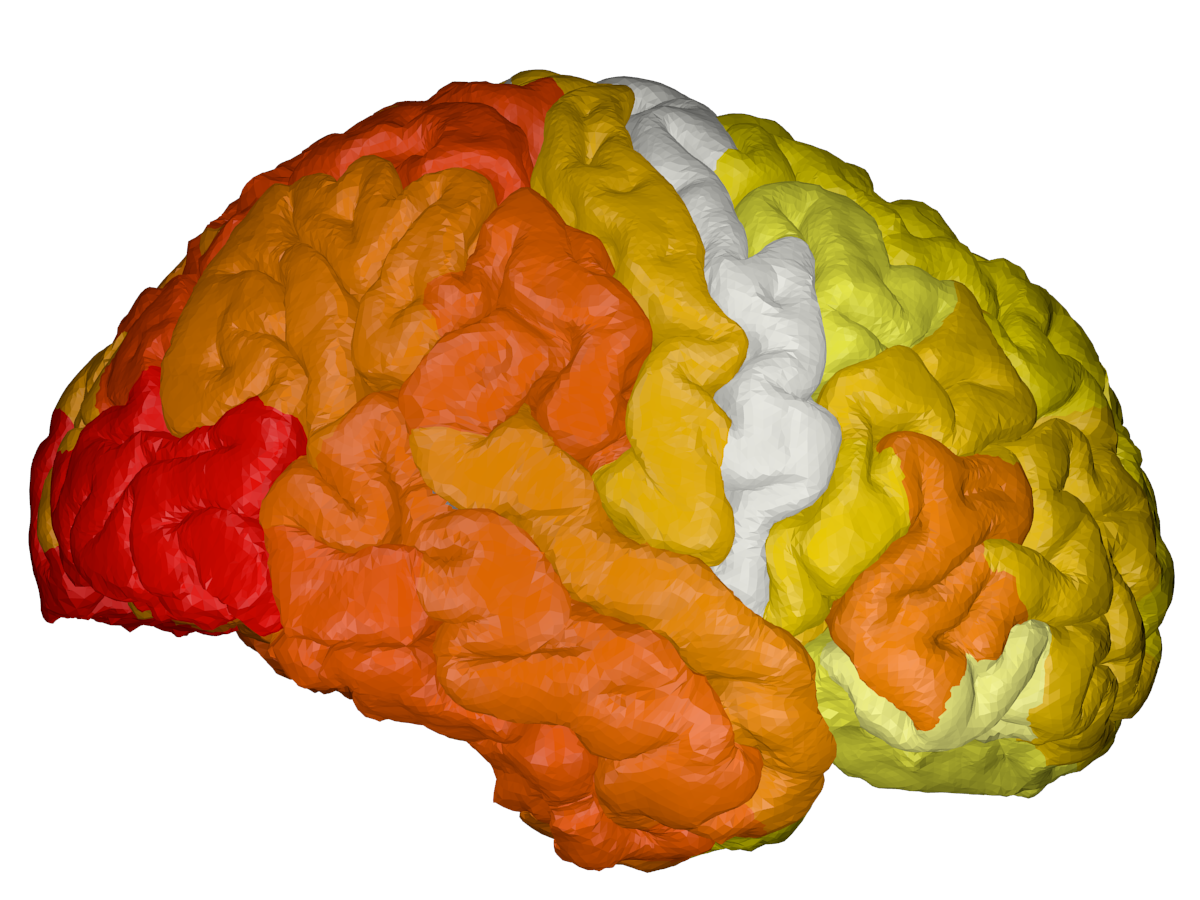}
 \includegraphics[height=3cm]{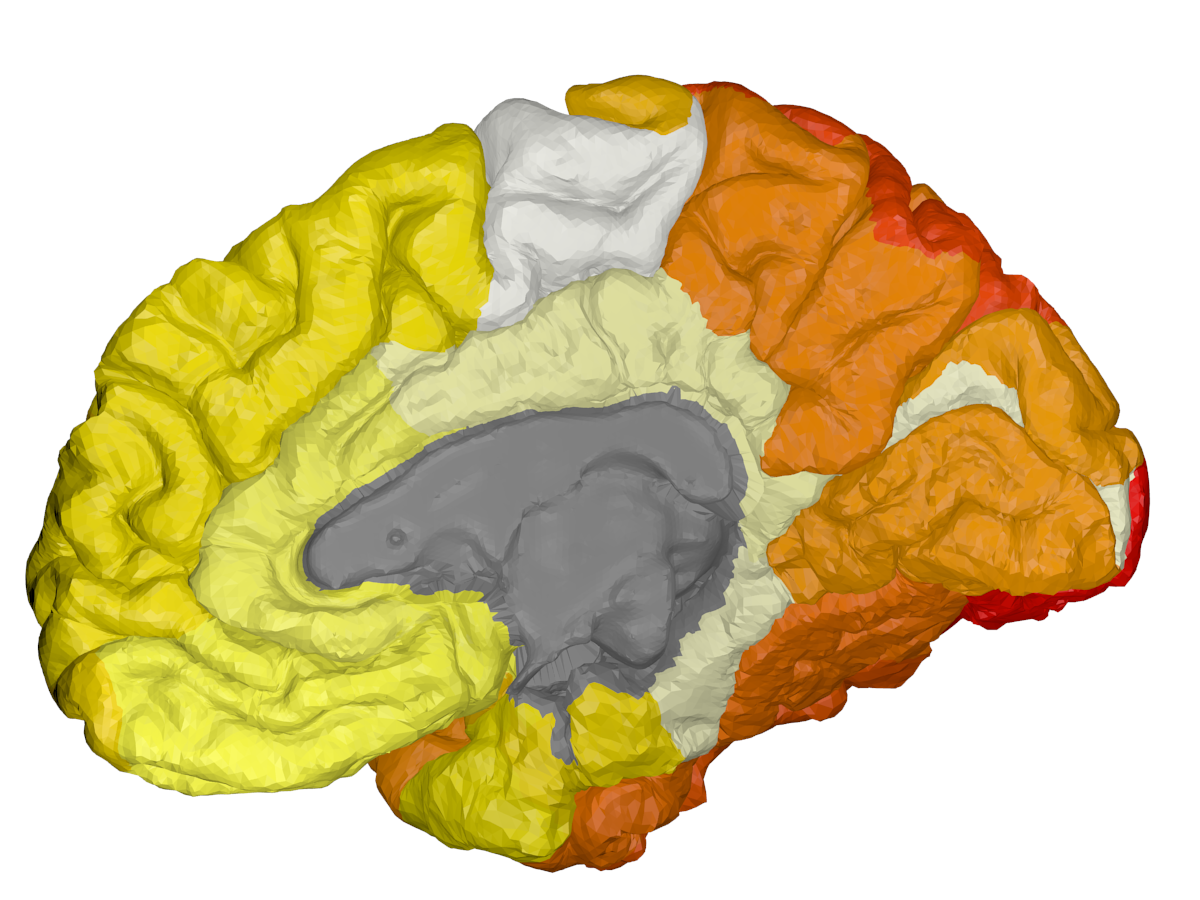}
 \includegraphics[height=3cm]{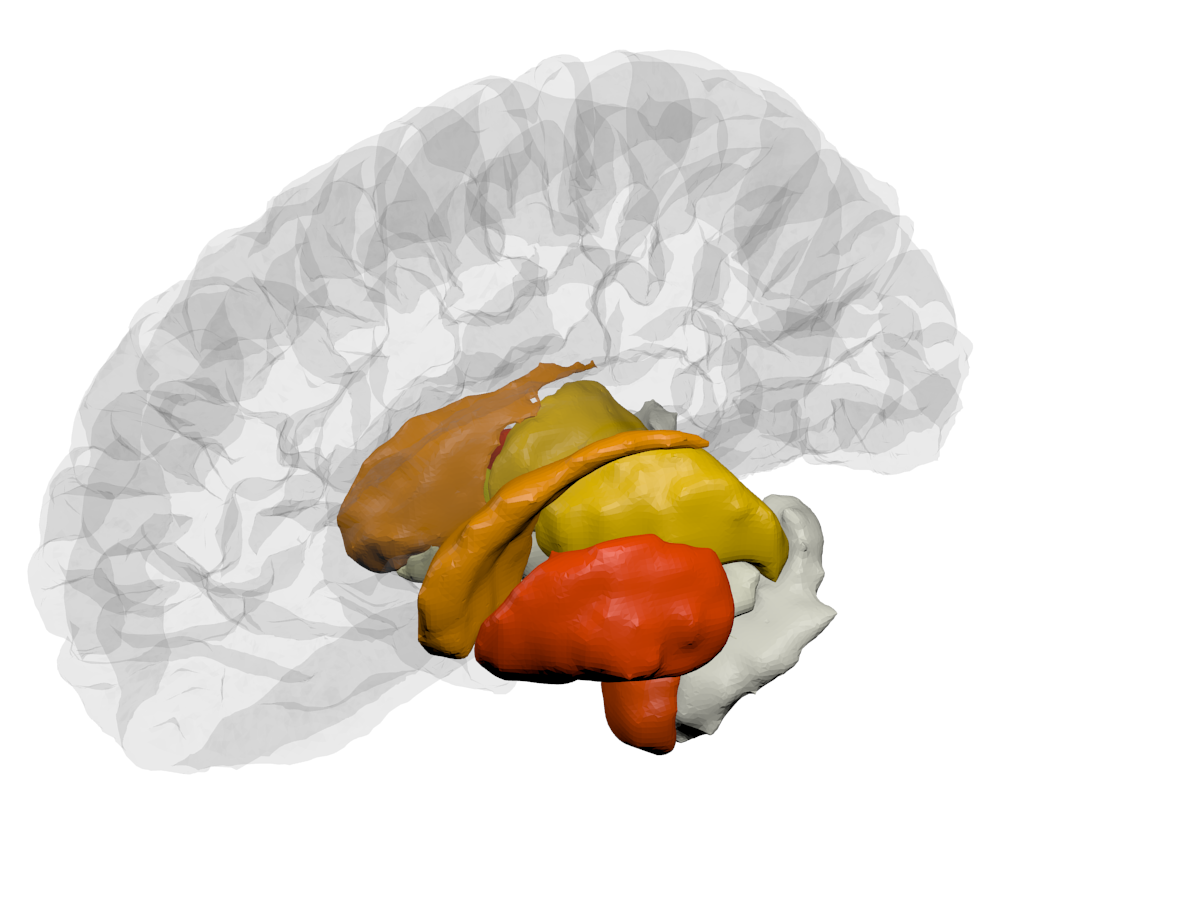}
} \\[\normalbaselineskip]
  \refstepcounter{figure}\normalfont\textbf{Figure~\thefigure: Some stuff about the teaser}
  \label{fig-teaser}
}
\makeatother

\usepackage[font=normalfont,labelfont=bf]{caption}

\makeatletter\ifdefined\Hy@StartlinkName
\def\addOneNestingLevelStartLink{%
  \gdef\Hy@StartlinkName##1##2{%
    \sbox0{\Hy@StartlinkNameOrig{##1}{##2}}\usebox0
    \global\let\Hy@StartlinkName\Hy@StartlinkNameOrig%
  }%
}
\def\addOneNestingLevelEndLink{%
  \gdef\pdfendlink{%
    \sbox0{\pdfendlinkOrig}\usebox0%
    \global\let\pdfendlink\pdfendlinkOrig%
  }%
}
\let\Hy@StartlinkNameOrig\Hy@StartlinkName
\let\pdfendlinkOrig\pdfendlink
\else
\let\addOneNestingLevelStartLink\relax
\let\addOneNestingLevelEndLink\relax
\fi
\makeatother

\usepackage{etoolbox}
\patchcmd{\emailauthor}{(#2)}{}{}{}
\patchcmd{\urlauthor}{(#2)}{}{}{}

\journal{}

\makeatletter
\def\ps@pprintTitle{%
 \let\@oddhead\@empty
 \let\@evenhead\@empty
 \def\@oddfoot{}%
 \let\@evenfoot\@oddfoot
 }
\makeatother

\usepackage{xcolor}

\usepackage{hyperref}
\definecolor{links}{HTML}{01368e}

\makeatletter
\hypersetup{colorlinks=true}
\AtBeginDocument{\@ifpackageloaded{hyperref}
  {\def\@linkcolor{links}
   \def\@anchorcolor{links}
   \def\@citecolor{links}
   \def\@filecolor{links}
   \def\@urlcolor{links}
   \def\@menucolor{links}
   \def\@pagecolor{links}
\begingroup
  \@makeother\`%
  \@makeother\=%
  \edef\x{%
    \edef\noexpand\x{%
      \endgroup
      \noexpand\toks@{%
        \catcode 96=\noexpand\the\catcode`\noexpand\`\relax
        \catcode 61=\noexpand\the\catcode`\noexpand\=\relax
      }%
    }%
    \noexpand\x
  }%
\x
\@makeother\`
\@makeother\=
}{}}
\makeatother

\usepackage{tikz}

\usepackage{filecontents}

\usepackage[round]{natbib}
\begin{document}

\newcommand{\frontFig}{
\begin{center}
\centering
\normalfont
\large

\begin{tikzpicture}
    \colorlet{redhsb}[hsb]{red}%
    \colorlet{yellowhsb}[hsb]{yellow}%
    \colorlet{orangehsb}[hsb]{orange}%
    \colorlet{whitehsb}[hsb]{white}%
    \shade[left color=white,right color=yellow] (-7,-0.5) rectangle (-5,-1); % 
    \shade[left color=yellow,right color=orange] (-5.02,-0.5) rectangle (-3,-1);
    \shade[left color=orange,right color=red] (-3.02,-0.5) rectangle (-1,-1); % 
    \node (G) at (-4.2,0) {User-defined color gradient};
    \draw (-7,-0.5) -- (-7,-1);\node[inner sep=0] (corr_text) at (-7,-1.3) {0};
    \draw (-5,-0.5) -- (-5,-1);\node[inner sep=0] (corr_text) at (-5,-1.3) {1};
    \draw (-3,-0.5) -- (-3,-1);\node[inner sep=0] (corr_text) at (-3,-1.3) {2};
    \draw (-1,-0.5) -- (-1,-1);\node[inner sep=0] (corr_text) at (-1,-1.3) {3};

    \node (A) at (-4.2,4) {INPUTS};
\node (A2) at (-4.2,2) {\fontsize{10}{12}\selectfont\begin{tabular}{c | c c c c}
 Biomarkers &  Hippocampus & Inferior & Superior & ...\\
(.csv file) &   & temporal & parietal & ...\\
  \hline
Brain 1 & 0.6 & 2.3 & 1.3 & .. \\
Brain 2 & 1.2 & 0.0 & 3.0 & .. \\
... & \multicolumn{4}{c}{...}\\
\end{tabular}};
 \node (B) at (5,4) {OUTPUTS};
 \node (C) at (5,2) {\includegraphics[height=2.3cm]{images/cortical-front_0.png}
\includegraphics[height=2.3cm]{images/cortical-back_0.png}
\includegraphics[height=2.3cm]{images/subcortical_0.png}
};
 \node (C2) at (C.north) {Brain 1};

    \node (D) at (5,-1) {\includegraphics[height=2.3cm]{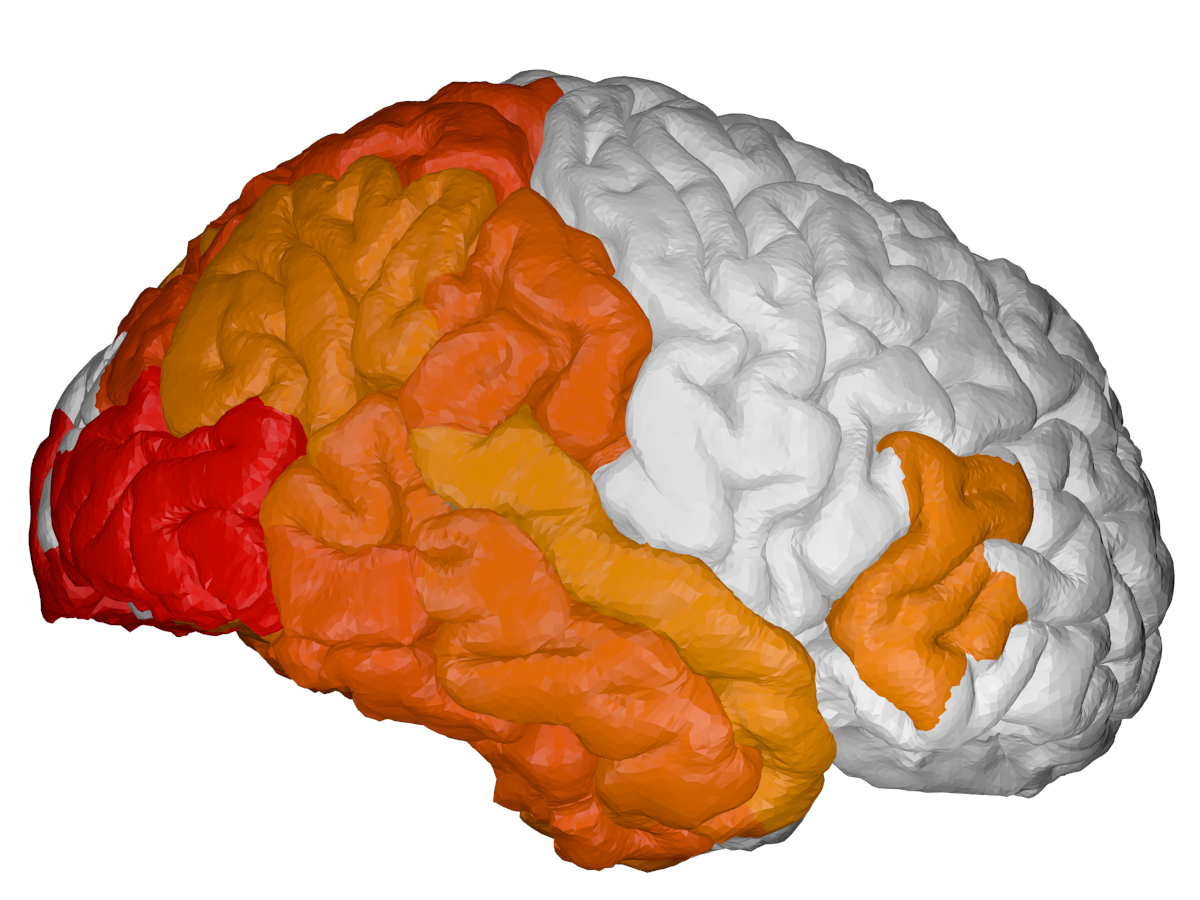}
\includegraphics[height=2.3cm]{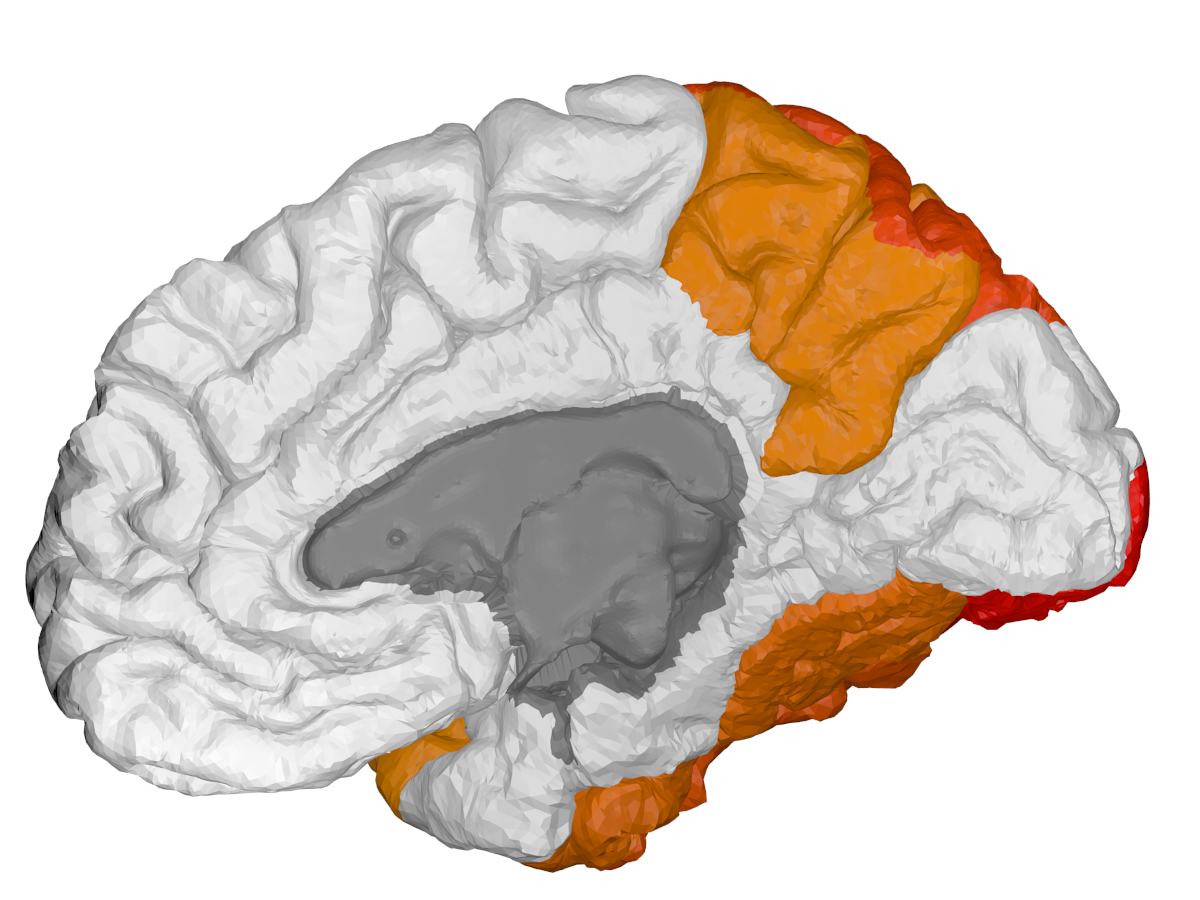}
\includegraphics[height=2.3cm]{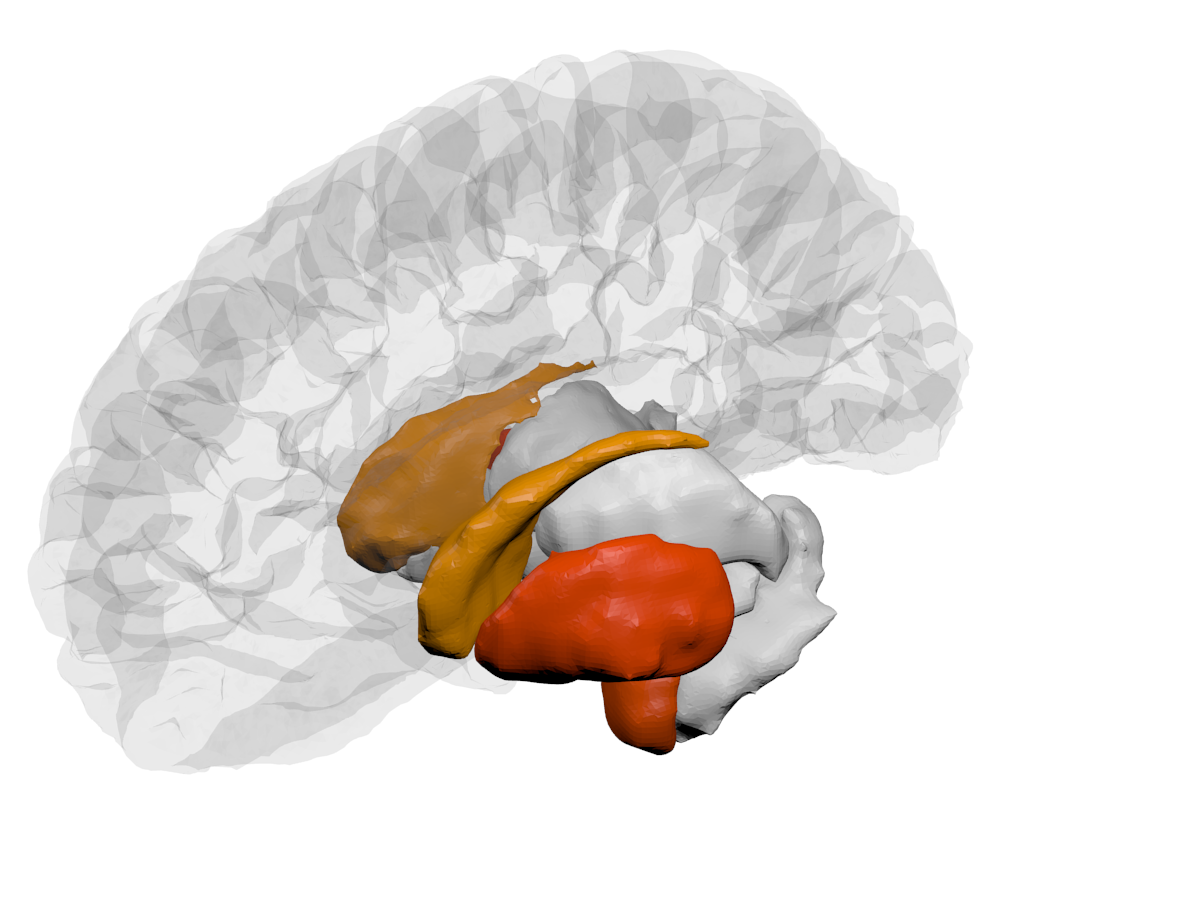}};
 \node (D2) at (D.north) {Brain 2};

    \draw[thick,->,shorten >= -2pt] (-2.7,4) -- (3,4) node[midway,above] {BrainPainter};
  \end{tikzpicture}
\captionof{figure}{Given a .csv file with region-of-interest (ROI) biomarkers and a user-defined color gradient, BrainPainter can automatically generate brain images with the cortical surface (left and middle) as well as with subcortical structures (right). The input .csv file can have multiple rows, one for each set of output images. The color gradient is a list of RGB colours given by the user. Final colours are interpolated using the numbers from the input .csv file based on the color gradient -- e.g. if the hippocampus has a value of 1.2, its final color will be an interpolation of colors 1 (yellow) and 2 (orange) from the gradient.}
\label{fig:mainDiag}
\end{center}
\vspace{-3em}
}

\begin{frontmatter}

\title{\textbf{BrainPainter: A software for the visualisation of brain structures, biomarkers and associated pathological processes}}

\ead{razvan@csail.mit.edu}
% \ead[url]{https://github.com/mrazvan22/brain-coloring}

\address[mit]{Computer Science and Artificial Intelligence Laboratory, Massachusetts Institute of Technology, Cambridge, USA, MA 02139}
\address[ucl]{Centre for Medical Image Computing, University College London, Gower Street, London, United Kingdom, WC1E 6BT}
\address[ion]{Queen Square MS Centre, UCL Institute of Neurology, UK
\frontFig
}

\cortext[aaa]{Joint senior authors with equal contribution}

\author[mit,ucl]{R\u{a}zvan V. Marinescu}
\author[ucl,ion]{Arman Eshaghi}
\author[ucl]{Daniel C. Alexander\corref{aaa}}
% \fntext[fn1]{* Joint senior authors}
\author[mit]{Polina Golland\corref{aaa}}
% \fntext[fn2]{*}

\begin{abstract}
We present BrainPainter, a software that automatically generates images of highlighted brain structures given a list of numbers corresponding to the output colours of each region. Compared to existing visualisation software (i.e. Freesurfer, SPM, 3D Slicer), BrainPainter has three key advantages: (1) it does not require the input data to be in a specialised format, allowing BrainPainter to be used in combination with any neuroimaging analysis tools, (2) it can visualise both cortical and subcortical structures and (3) it can be used to generate movies showing dynamic processes, e.g. propagation of pathology on the brain. We highlight three use cases where BrainPainter was used in existing neuroimaging studies: (1) visualisation of the degree of atrophy through interpolation along a user-defined gradient of colours, (2) visualisation of the progression of pathology in Alzheimer's disease as well as (3) visualisation of pathology in subcortical regions in Huntington's disease. Moreover, through the design of BrainPainter we demonstrate the possibility of using a powerful 3D computer graphics engine such as Blender to generate brain visualisations for the neuroscience community. Blender's capabilities, e.g. particle simulations, motion graphics, UV unwrapping, raster graphics editing, raytracing and illumination effects, open a wealth of possibilities for brain visualisation not available in current neuroimaging software. BrainPainter is customisable, easy to use, and can run straight from the web browser: \url{https://brainpainter.csail.mit.edu}, as well as from source-code packaged in a docker container: \url{https://github.com/mrazvan22/brain-coloring}. It can be used to visualise biomarker data from any brain imaging modality, or simply to highlight a particular brain structure for e.g. anatomy courses.

\end{abstract}

% \begin{keyword}
% Brain visualisation \sep
% Neurodegenerative diseases
% \end{keyword}

\end{frontmatter}

%% \linenumbers

%% main text

% \FloatBarrier

% \section*{Abstract}
% % \label{intro}

\section{Introduction}
\label{intro}

% \frontFig

% diagram showing the aim: input numbers and output images

Efficient visualisation of brain structure, function and pathology is crucial for understanding the mechanisms underlying neurodegenerative diseases and eases the interpretation of results in neuroimaging studies. This is especially important in populations studies, where two or more populations are compared for group differences in biomarkers derived from e.g. Magnetic Resonance Imaging, Positron Emission   Tomography (PET) or Computer Tomography (CT). The results are best visualised as brain images, where regions-of-interest (ROIs) are highlighted based on the magnitude of the difference between the two groups. These visualisation are generally done by the same software that performs the registration, segmentation and statistical analysis. However, for traumatic brain injury or less common neurodegenerative diseases such as Parkinson's disease and Multiple Sclerosis, visualisations of statistical results is sometimes not performed due to the inability to register images to a common template or lack of robust registration software. Therefore, many studies such as (\cite{coughlin2015neuroinflammation,schoonheim2012subcortical}) only report differences between patients and controls in tables or as box plots. There is therefore a lack of visualisation tools that can highlight neuroimaging findings for these complex diseases.

When registration to a common population template is possible, e.g. in Alzheimer's disease (AD), excellent 3D visualisation software exists  which allows interactive visualisation of population differences -- e.g. 3D Slicer (\cite{pieper20043d}), Freesurfer (\cite{fischl2012freesurfer}) or SPM (\cite{penny2011statistical}). However, they have several inherent limitations. First, such software -- e.g. Freesurfer -- generally require inputs in their proprietary data format, which is usually difficult and time-consuming to create without using their pipeline. While creating these proprietary data formats is necessary when users need to display voxelwise visualisations, often users only need to highlight entire ROIs -- in this simpler case the user could only provide a list of RGB colors for each ROI in a csv file, removing the need to create input data in a specialised format. Another limitation of existing visualisation software is their difficulty in highlighting complex patterns of pathology in a single slice of a 3D volumetric image. To overcome this, some authors show multiple slices (sometimes up to 8 slices, e.g. \cite{migliaccio2015mapping}), although this takes too much space on the academic paper being published. While Freesurfer solves this problem using a cortical surface-based representation that captures most of the complexity of pathology patterns in a single image, this surface representation is not supported for subcortical structures such as the hippocampus. Third, current visualisation software cannot be easily used to generate e.g. a movie showing a dynamic process, e.g. propagation of pathology within the human brain, as most of them have been intended for interactive visualisation and have no application programming interface (API) that allows automatic generation of hundreds of images using pre-defined settings. 

We present BrainPainter, a software for easy visualisation of structures, pathology and biomarkers in the brain. As opposed to previous visualisation software, the input data is very simple: a generic .csv file containing numbers for each ROI, each number mapping to a different colour to be assigned to that ROI -- such a simple input allows BrainPainter to be used in conjunction with any other neuroimaging analysis software. Secondly, BrainPainter can visualise both cortical and subcortical structures using a surface representation, removing the need to show multiple slices of the same 3D scan. Third, the images are generated automatically from pre-defined view-points, and can be easily used to create a movie showing e.g. the propagation of pathology, without the need to write any extra software code or interface with an API. Written in Python, BrainPainter is open-source and can be extended in many ways. In this paper we describe the capabilities of the software, and showcase three use cases where it was used in neuroimaging studies.

\FloatBarrier
\section{Design}
\label{design}

% diagram showing the Design. what kind of atlases can it take, type of colouring, blender integration

\begin{figure*}[htp]
\centering
 \includegraphics[width=0.8\textwidth]{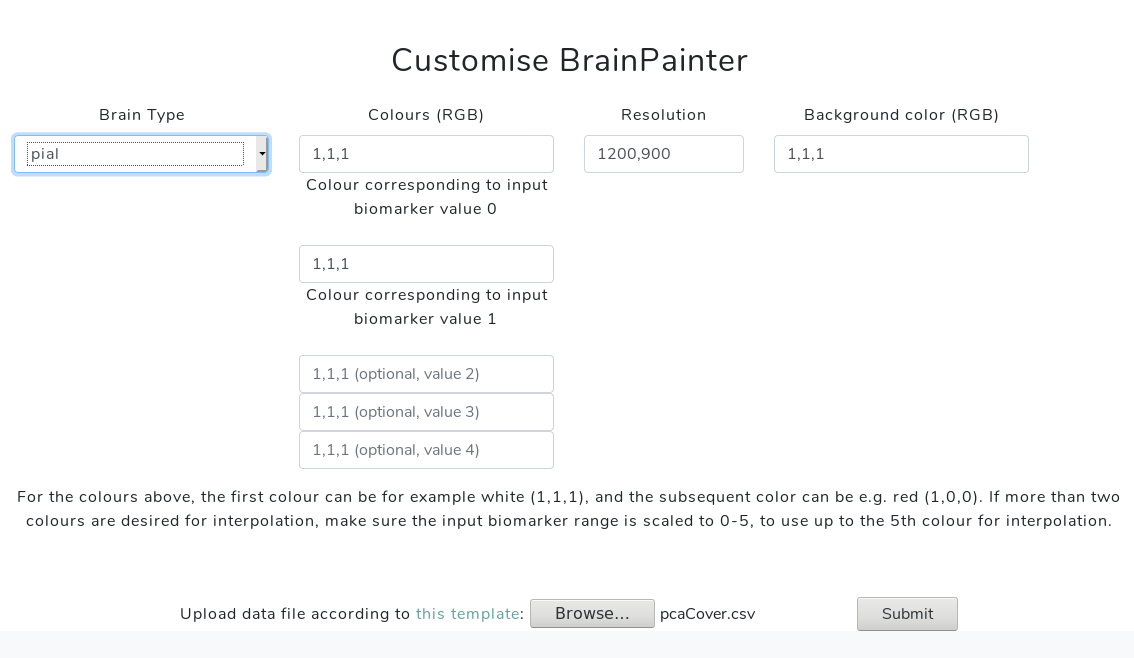}
 \caption{BrainPainter website interface at \url{http://brainpainter.csail.mit.edu}, showing how it can be easily customised. Here, the user selects the brain type, colours and resolution, and finally uploads the input .csv file with ROI biomarkers. The server then generates the output images, which can be downloaded by the user. More customisation features will be added in future versions.}
  \label{fig:website}
\end{figure*}

% how does BrianPainter work: takes input csv file, generates images. 
BrainPainter has a very simple yet effective workflow. Given an input csv file with biomarkers for each region, it produces high-quality visualisations of cortical and subcortical structures. For this, it uses Blender as a rendering engine, and loads 3D meshes from a template brain (one 3D mesh for each ROI), which are then coloured according to the input numbers. Instead of providing a list of RGB colours for each ROI, we choose a simpler interface of providing one number for each ROI which maps to an RGB color using a user-defined color gradient. For example, the gradient can range from white $\rightarrow$ yellow $\rightarrow$ orange $\rightarrow$ red, as in the example from Fig. \ref{fig:mainDiag}. In this case, the input numbers for each ROI need to be in the range [0,3], where a value of 1.3 would interpolate between colour 1 (yellow) and color 2 (orange).

% use of blender as rednering engine
BrianPainter uses open-source software Blender as the rendering engine. We chose Blender for three reasons. First, it is open-source, allowing us to distribute it already integrated with BrainPainter, thus requiring no further installation. Secondly, Blender is a powerful 3D graphics software, which allowed us to create realistic lightning conditions and handle transparency required for the glass-brain. Third, it also supports creating movies of complex temporal processes such as pathology spread along the brain. The software also supports a variety of object formats for the brain template, including the popular .obj mesh format. As BrainPainter is written in Python, it allows interfacing with any Blender function.

% ATLASES: DK
The software is able to colour and visualise regions belonging to a pre-defined atlas. Currently, we support three widely-used atlases: (i) the Desikan-Killiany (DK) atlas (\cite{desikan2006automated}), (ii) the Destrieux atlas (\cite{destrieux2010automatic}) and (iii) the Tourville atlas (\cite{klein2012101}). However, a custom atlas can also be used by mapping those regions to any of the three atlases currently supported, through the modification a simple mapping in the main configuration file. \footnote{\url{https://github.com/mrazvan22/brain-coloring/blob/master/config.py}}

\section{Customisation}
\label{customisation}

% customisation

BrainPainter can be easily customised in several ways, as shown in Fig \ref{fig:website}. First of all, the colours assigned to each region can be changed by modifying both the control points of the color gradient and the input numbers selecting colors along the gradient. The background colour and image resolution can also be changed.  

\begin{figure*}[htp]
\centering
 \includegraphics[width=0.8\textwidth, trim=20 0 0 220, clip]{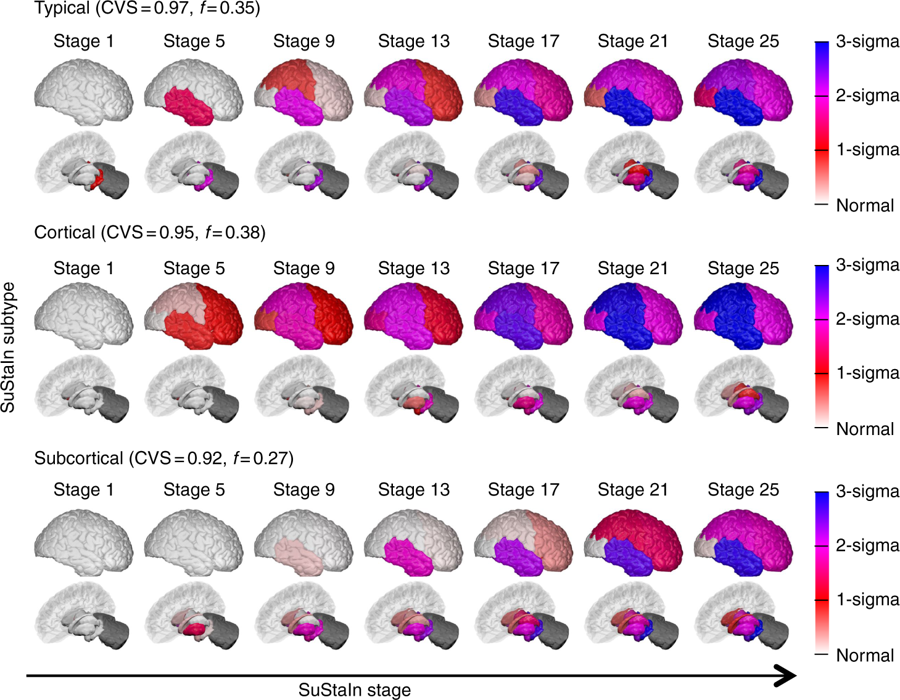}
 \caption{Demonstration of our software for showing the temporal progression of pathology, as a sequence of snapshot at different stages. Images used by
 \cite{young2018uncovering}.}
  \label{fig:youngProg}
\end{figure*}

The 3D structures being visualised can also be customised. We currently support three atlases (Desikan-Killiany, Destrieux and Tourville) as well as two types of brain surfaces: inflated, which is a brain surface that is smoothed out and where no gyri appear, and also pial, the standard brain surface with gyri. The software allows one to remove some 3D structures -- for example, Fig. \ref{fig:peter} shows the subcortical structures with the cerebellum removed from the visualisation -- contrast this with Fig. \ref{fig:youngProg}. 
 
BrainPainter also support two types of surfaces, cortical and subcortical structures. For the cortical surface, we only show the left hemisphere (although the right hemisphere can also be added), and provide two default viewing angles (front and back). For the subcortical structures, we show them for both hemispheres and also show the right hemisphere as a glass brain, for reference.

More complex settings such as the viewing angle and luminosity can also be customised, but currently require minor modifications to the source code. In the future, we plan to enable these customisations from the main configuration file.

\begin{figure}
 \includegraphics[width=0.49\textwidth]{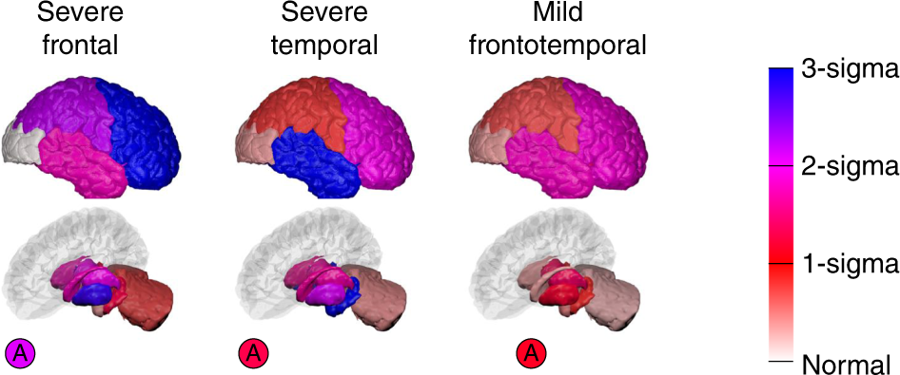}
 \caption{Demonstration of BrainPainter for showing extent of pathology on a vertical bar, where colours towards blue show increased severity. Source: \cite{young2018uncovering}.}
 \label{fig:youngDegree}
\end{figure}

\section{First use case: Visualising the degree of pathology}
\label{degree}

In the first use case, we want to visualise the degree of pathology in Alzheimer's disease. During the progression of Alzheimer's disease, some regions of the brain such as the hippocampus and temporal lobes will be more affected compared to other regions of the brain such as the occipital lobe. Visualisation of pathology in AD is important in order to understand its underlying mechanisms and generate new hypotheses. 

The notion of pathology here is abstract, and can refer to atrophy as measured by volume loss or cortical thinning, white matter degradation as measured by diffusion tensor imaging (DTI) changes in fractional anisotropy (FA), or the level of abnormal conformations of proteins such as amyloid-beta or tau as measured by Positron Emission Tomography. However, BrainPainter is agnostic to the meaning of these biomarkers and can be used with any imaging modality, including markers derived from several modalities together. 

Fig. \ref{fig:youngDegree} shows an application of BrainPainter by (\cite{young2018uncovering}) to highlight the degree of atrophy in Alzheimer's disease. Regions with no atrophy are coloured in white, while regions with severe atrophy are coloured in blue. The gradient on the right shows, for every color, the number of standard deviations away from controls.

\section{Second use case: Visualising the temporal progression of neurodegenerative diseases}
\label{progression2}

In the second use case, we would like to visualise the temporal progression of Alzheimer's disease (AD). Alzheimer's disease is characterised by a slow, continuous progression -- while it's mechanisms are still not fully understood, it is currently believed that initial abnormalities in the amyloid and tau proteins cause a cascade of events that eventually lead to axonal degradation, neural death and cognitive decline (\cite{mudher2002alzheimer}). Therefore, being able to visualise the progression of these events, including their timing and speed, is crucial for understanding the mechanisms of Alzheimer's disease.

Fig. \ref{fig:youngProg}, reproduced and adapted from \cite{young2018uncovering}, demonstrates the ability of BrainPainter to visualise the evolution of atrophy in two subtypes of Alzheimer's disease -- \emph{cortical} and \emph{subcortical} -- characterised by prominent atrophy in the cortical and subcortical regions respectively. This study done by \cite{young2018uncovering} used data from the Alzheimer's disease Neuroimaging Initiative to disentangle the heterogeneity of AD into subtypes with different progression. Here, visualisations provided by BrainPainter were able to characterise not only the degree of atrophy in each region (white/red to blue colors), but also the timing of atrophy events. For example, even in the \emph{cortical} subtype, the hippocampus becomes affected by stage 13, while similarly, in the \emph{subcortical} subtype the temporal lobe becomes affected by stage 13.

\section{Third use case: Visualising pathology in subcortical structures}
\label{progression3}

The ability to visualise subcortical structures is crucial for neurodegenerative diseases that cause damage to these regions. Apart from Alzheimer's disease, Huntington's disease (HD) is also known for targetting subcortical regions (\cite{douaud2009vivo,wijeratne2018image}). The neurodegeneration in HD is believed to begin in the striatum and pallidum, and later followed by other subcortical and cortical regions (\cite{douaud2009vivo}).

\begin{figure}[htp]
\centering
% \subfigure{Stage 0\includegraphics[width=0.2\textwidth]{images/ebmhd_pngs/subcortical_stage0.png}}
\subfloat[Stage 0]{
\includegraphics[width=0.2\textwidth,trim=70 0 70 0, clip]{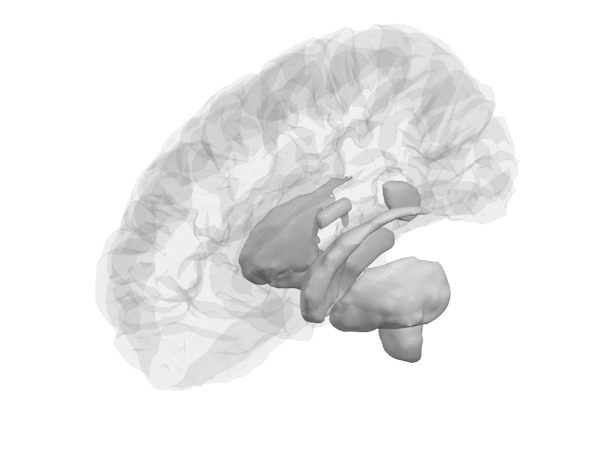}}
\subfloat[Stage 3]{
\includegraphics[width=0.2\textwidth,trim=70 0 70 0, clip]{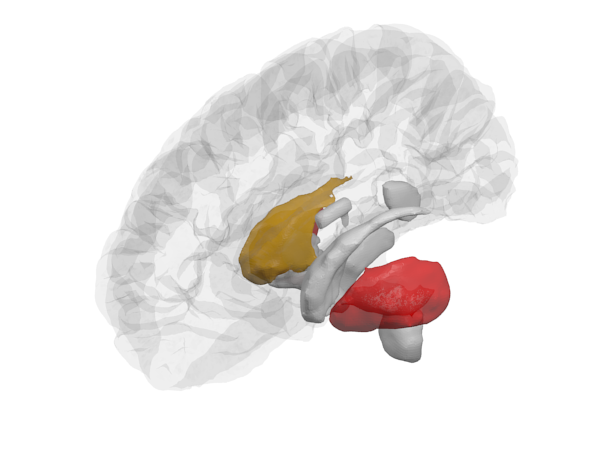}}

\subfloat[Stage 7]{
\includegraphics[width=0.2\textwidth,trim=70 0 70 0, clip]{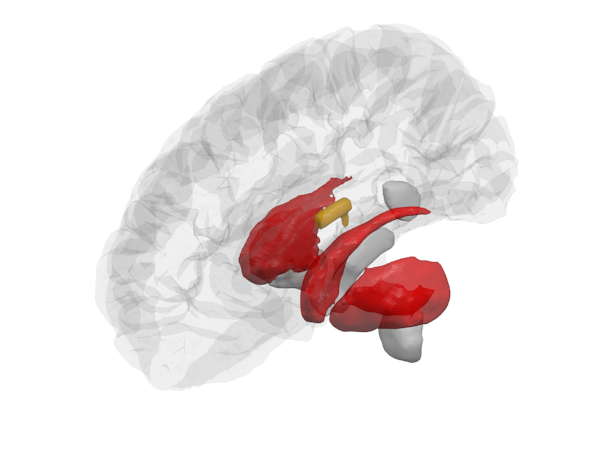}}
\subfloat[Stage 10]{
\includegraphics[width=0.2\textwidth,trim=70 0 70 0, clip]{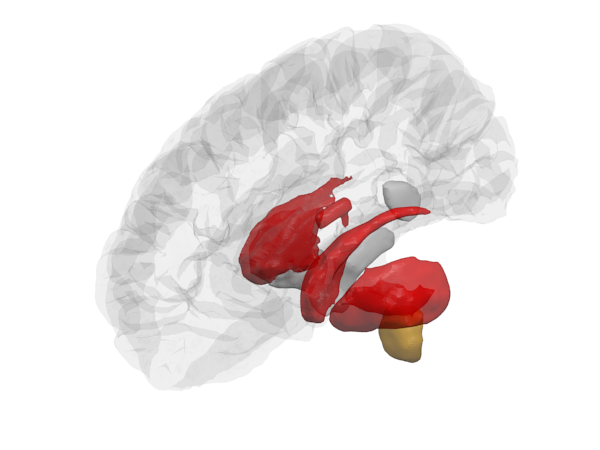}}
 \caption{Progression of pathology in subcortical regions within a glass brain, using images generated with our method. Images used by  
 \cite{wijeratne2018image}.}
   \label{fig:peter}
\end{figure}

Fig. \ref{fig:peter}, reproduced and adapted with permission from \cite{wijeratne2018image}, shows visualisations generated by BrainPainter of atrophy progression in subcortical areas, for Huntington's disease. The images show early involvement of the putamen, caudate and pallidum in the progression of Huntington's disease, and demonstrate the potential of BrainPainter in visualising pathology dynamics in subcortical regions using parsimonious glass-brain images.

\section{Conclusion}

We presented BrainPainter, an open-source software that can be used to visualise structures, biomarkers and pathologies in the human brain. The visualisations generated by BrainPainter can be used to significantly enhance the interpretation of neuroimaging research and can be easily embedded by researchers into scientific articles. While not demonstrated here, BrainPainter can also easily generate movies showing dynamic processes, e.g. propagation of brain pathology. 

Our software has several limitations that can be addressed in future versions. First, it can currently only highlight entire regions-of-interest from an atlas. However, this was a design choice, as it removes the need for users to create specialised input files with voxelwise measurements, thus increasing usability. Nevertheless, in future versions we might add the ability to highlight fine-grained patterns of pathology. Yet another limitation of BrainPainter is that it cannot visualise more complex structures such as white-matter tracts, although we plan to add such functionality in future releases. 

The use of the powerful Blender engine opens numerous avenues not possible with current neuroimaging software: motion graphics can be used to generate realistic movies showing e.g. the evolution of biomarkers, particle simulations can be used to visualise toxic proteins accumulating in certain regions, soft-body simulations can be used to model brain deformations due to head trauma, while camera-based rendering allows the creation of educational videos. 

\FloatBarrier
\section{Acknowledgements}

RVM was supported by the NIH grants NIBIB NAC P41EB015902 and NINDS R01NS086905, as well as the EPSRC Centre For Doctoral Training in Medical Imaging with grant EP/L016478/1. AE received a McDonald Fellowship from the Multiple Sclerosis International Federation (MSIF, www.msif.org), and the ECTRIMS -- MAGNIMS Fellowship. DCA was supported by EuroPOND, which is an EU Horizon 2020 project, and by EPSRC grants J020990, M006093 and M020533. PG was supported by NIH grants NIBIB NAC P41EB015902 and NINDS R01NS086905. 

We are also particularly grateful to Anderson Winkler for creating the 3D brain templates for all three atlases, which are used in this work.\footnote{https://brainder.org/research/brain-for-blender/}

\section{References}

\bibliographystyle{plainnat}
\bibliography{bibliography}

\begin{thebibliography}{13}
\providecommand{\natexlab}[1]{#1}
\providecommand{\url}[1]{\texttt{#1}}
\expandafter\ifx\csname urlstyle\endcsname\relax
  \providecommand{\doi}[1]{doi: #1}\else
  \providecommand{\doi}{doi: \begingroup \urlstyle{rm}\Url}\fi

\bibitem[Coughlin et~al.(2015)Coughlin, Wang, Munro, Ma, Yue, Chen, Airan, Kim,
  Adams, Garcia, et~al.]{coughlin2015neuroinflammation}
Jennifer~M Coughlin, Yuchuan Wang, Cynthia~A Munro, Shuangchao Ma, Chen Yue,
  Shaojie Chen, Raag Airan, Pearl~K Kim, Ashley~V Adams, Cinthya Garcia, et~al.
\newblock Neuroinflammation and brain atrophy in former {NFL} players: an in
  vivo multimodal imaging pilot study.
\newblock \emph{Neurobiology of disease}, 74:\penalty0 58--65, 2015.

\bibitem[Desikan et~al.(2006)Desikan, S{\'e}gonne, Fischl, Quinn, Dickerson,
  Blacker, Buckner, Dale, Maguire, Hyman, et~al.]{desikan2006automated}
Rahul~S Desikan, Florent S{\'e}gonne, Bruce Fischl, Brian~T Quinn, Bradford~C
  Dickerson, Deborah Blacker, Randy~L Buckner, Anders~M Dale, R~Paul Maguire,
  Bradley~T Hyman, et~al.
\newblock An automated labeling system for subdividing the human cerebral
  cortex on {MRI} scans into gyral based regions of interest.
\newblock \emph{Neuroimage}, 31\penalty0 (3):\penalty0 968--980, 2006.

\bibitem[Destrieux et~al.(2010)Destrieux, Fischl, Dale, and
  Halgren]{destrieux2010automatic}
Christophe Destrieux, Bruce Fischl, Anders Dale, and Eric Halgren.
\newblock Automatic parcellation of human cortical gyri and sulci using
  standard anatomical nomenclature.
\newblock \emph{Neuroimage}, 53\penalty0 (1):\penalty0 1--15, 2010.

\bibitem[Douaud et~al.(2009)Douaud, Behrens, Poupon, Cointepas, Jbabdi, Gaura,
  Golestani, Krystkowiak, Verny, Damier, et~al.]{douaud2009vivo}
Gwena{\"e}lle Douaud, Timothy~E Behrens, Cyril Poupon, Yann Cointepas, Sa{\^a}d
  Jbabdi, V{\'e}ronique Gaura, Narly Golestani, Pierre Krystkowiak, Christophe
  Verny, Philippe Damier, et~al.
\newblock In vivo evidence for the selective subcortical degeneration in
  {Huntington}'s disease.
\newblock \emph{Neuroimage}, 46\penalty0 (4):\penalty0 958--966, 2009.

\bibitem[Fischl(2012)]{fischl2012freesurfer}
Bruce Fischl.
\newblock Freesurfer.
\newblock \emph{Neuroimage}, 62\penalty0 (2):\penalty0 774--781, 2012.

\bibitem[Klein and Tourville(2012)]{klein2012101}
Arno Klein and Jason Tourville.
\newblock 101 labeled brain images and a consistent human cortical labeling
  protocol.
\newblock \emph{Frontiers in neuroscience}, 6:\penalty0 171, 2012.

\bibitem[Migliaccio et~al.(2015)Migliaccio, Agosta, Possin, Canu, Filippi,
  Rabinovici, Rosen, Miller, and Gorno-Tempini]{migliaccio2015mapping}
Raffaella Migliaccio, Federica Agosta, Katherine~L Possin, Elisa Canu, Massimo
  Filippi, Gil~D Rabinovici, Howard~J Rosen, Bruce~L Miller, and Maria~Luisa
  Gorno-Tempini.
\newblock Mapping the progression of atrophy in early-and late-onset
  alzheimer’s disease.
\newblock \emph{Journal of Alzheimer's Disease}, 46\penalty0 (2):\penalty0
  351--364, 2015.

\bibitem[Mudher and Lovestone(2002)]{mudher2002alzheimer}
Amritpal Mudher and Simon Lovestone.
\newblock Alzheimer's disease--do tauists and baptists finally shake hands?
\newblock \emph{Trends in neurosciences}, 25\penalty0 (1):\penalty0 22--26,
  2002.

\bibitem[Penny et~al.(2011)Penny, Friston, Ashburner, Kiebel, and
  Nichols]{penny2011statistical}
William~D Penny, Karl~J Friston, John~T Ashburner, Stefan~J Kiebel, and
  Thomas~E Nichols.
\newblock \emph{Statistical parametric mapping: the analysis of functional
  brain images}.
\newblock Elsevier, 2011.

\bibitem[Pieper et~al.(2004)Pieper, Halle, and Kikinis]{pieper20043d}
Steve Pieper, Michael Halle, and Ron Kikinis.
\newblock 3d slicer.
\newblock In \emph{2004 2nd IEEE international symposium on biomedical imaging:
  nano to macro (IEEE Cat No. 04EX821)}, pages 632--635. IEEE, 2004.

\bibitem[Schoonheim et~al.(2012)Schoonheim, Popescu, Lopes, Wiebenga, Vrenken,
  Douw, Polman, Geurts, and Barkhof]{schoonheim2012subcortical}
Menno~M Schoonheim, Veronica Popescu, Fernanda C~Rueda Lopes, Oliver~T
  Wiebenga, Hugo Vrenken, Linda Douw, Chris~H Polman, Jeroen~JG Geurts, and
  Frederik Barkhof.
\newblock Subcortical atrophy and cognition: sex effects in multiple sclerosis.
\newblock \emph{Neurology}, 79\penalty0 (17):\penalty0 1754--1761, 2012.

\bibitem[Wijeratne et~al.(2018)Wijeratne, Young, Oxtoby, Marinescu, Firth,
  Johnson, Mohan, Sampaio, Scahill, Tabrizi, et~al.]{wijeratne2018image}
Peter~A Wijeratne, Alexandra~L Young, Neil~P Oxtoby, Razvan~V Marinescu,
  Nicholas~C Firth, Eileanoir~B Johnson, Amrita Mohan, Cristina Sampaio,
  Rachael~I Scahill, Sarah~J Tabrizi, et~al.
\newblock An image-based model of brain volume biomarker changes in
  {Huntington}'s disease.
\newblock \emph{Annals of clinical and translational neurology}, 5\penalty0
  (5):\penalty0 570--582, 2018.

\bibitem[Young et~al.(2018)Young, Marinescu, Oxtoby, Bocchetta, Yong, Firth,
  Cash, Thomas, Dick, Cardoso, et~al.]{young2018uncovering}
Alexandra~L Young, Razvan~V Marinescu, Neil~P Oxtoby, Martina Bocchetta, Keir
  Yong, Nicholas~C Firth, David~M Cash, David~L Thomas, Katrina~M Dick, Jorge
  Cardoso, et~al.
\newblock Uncovering the heterogeneity and temporal complexity of
  neurodegenerative diseases with {Subtype} and {Stage} {Inference}.
\newblock \emph{Nature communications}, 9\penalty0 (1):\penalty0 4273, 2018.

\end{thebibliography}

\end{document}